\documentclass[usenatbib]{mnras}
\usepackage{natbibmnfix, graphicx, times, amsmath, epsfig, amssymb}
\usepackage{verbatim}
\usepackage{color}
\usepackage{mathptmx}
\usepackage{txfonts}
\usepackage[T1]{fontenc}
\usepackage{aecompl}

\def\simgt{\lower.5ex\hbox{\gtsima}} 
\def\simlt{\lower.5ex\hbox{\ltsima}} 
\def\gtsima{$\; \buildrel > \over \sim \;$} 
\def\ltsima{$\; \buildrel < \over \sim \;$}
\def\Msun{M_\odot}

\newcommand\lsim{\mathrel{\rlap{\lower4pt\hbox{\hskip1pt$\sim$}}
        \raise1pt\hbox{$<$}}}
\newcommand\gsim{\mathrel{\rlap{\lower4pt\hbox{\hskip1pt$\sim$}}
        \raise1pt\hbox{$>$}}}
\def\myputfigure#1#2#3#4#5%
{\vskip#5pt\makebox[0pt]{\hskip#2in
\includegraphics[width=#3\textwidth]{#1}}\vskip#4pt\hfill}

\title[The Spectral Evolution of the First Black Holes]
      {Shining in the Dark: the Spectral Evolution of the First Black Holes}
\author[F. Pacucci et al.]
{Fabio Pacucci$^1$ \thanks{fabio.pacucci@sns.it},
Andrea Ferrara$^{1,2}$, Marta Volonteri$^{3}$, Guillaume Dubus$^{3,4}$\\
$^1$Scuola Normale Superiore, Piazza dei Cavalieri, 7  56126 Pisa, Italy \\
$^2$Kavli Institute for the Physics and Mathematics of the Universe (WPI), Todai Institutes for Advanced Study, \\ 
\, the University of Tokyo 5-1-5 Kashiwanoha, Kashiwa, 277-8583, Japan \\
$^3$Sorbonne Universit\'{e}s, UPMC Univ Paris 06, CNRS, UMR 7095, Institut d'Astrophysique de Paris, F-75014 Paris, France \\
$^4$Universit\'{e} Grenoble Alpes, CNRS, UMR 5274, IPAG, F-38000 Grenoble, France\\
}
             
\date{submitted to MNRAS}
\pubyear{2015}

\begin{document}
\label{firstpage}
\pagerange{\pageref{firstpage}--\pageref{lastpage}}
\maketitle
             
\begin{abstract}
Massive Black Hole (MBH) seeds at redshift $z \gsim 10$ are now thought to be key ingredients to explain the presence of the super-massive ($10^{9-10} \, \mathrm{\Msun}$) black holes in place $ < 1 \, \mathrm{Gyr}$ after the Big Bang. Once formed, massive seeds grow and emit copious amounts of radiation by accreting the left-over halo gas; their spectrum can then provide crucial information on their evolution.  By combining radiation-hydrodynamic and spectral synthesis codes, we simulate the time-evolving spectrum emerging from the host halo of a MBH seed with initial mass $10^5 \, \mathrm{\Msun}$, assuming both standard Eddington-limited accretion, or slim accretion disks, appropriate for super-Eddington flows.
The emission occurs predominantly in the \textit{observed} infrared-submm ($1-1000 \, \mathrm{\mu m}$) and X-ray ($0.1 - 100 \, \mathrm{keV}$) bands.
Such signal should be easily detectable by JWST around $\sim 1 \, \mathrm{\mu m}$ up to $z \sim 25$,  and by ATHENA (between $0.1$ and $10 \, \mathrm{keV}$, up to $z \sim 15$). Ultra-deep X-ray surveys like the Chandra Deep Field South could have already detected these systems up to $z \sim 15$. Based on this, we provide an upper limit for the $z \gsim 6$ MBH mass density of $\rho_{\bullet}  \lsim 2.5 \times 10^{2} \, \mathrm{\Msun \, Mpc^{-3}}$ assuming standard Eddington-limited accretion.  If accretion occurs in the slim disk mode the limits are much weaker,  $\rho_{\bullet}  \lsim 7.6 \times 10^{3} \, \mathrm{\Msun \, Mpc^{-3}}$ in the most constraining case.
\end{abstract}


\setcounter{footnote}{1}
\newcounter{dummy}
\section{Introduction}
\label{sec:introduction}
The cosmic epoch in the redshift range $10 \lsim z \lsim 30$ was characterized by the formation of the first stars (Pop III) and of the first black holes (see \citealt{2011ARA&A..49..373B}, \citealt{Volonteri_2010}, \citealt{Volonteri_Bellovary_2012}, \citealt{Haiman_2013} for recent reviews).
Detecting these sources directly is at, or beyond, the sensitivity edge of current observatories. While recently \cite{Sobral_2015} have shown that a very luminous $\mathrm{Ly}\alpha$ emitter at $z \approx 6.6$ may be consistent with having a mixed composition of Pop III and second-generation (Pop II) stars, to date there are no confirmed observations of the first black holes, partly due to the uncertainty on their observational signatures. The next generation of observatories will most likely detect the first glimpses of light in the Universe, both in the electromagnetic spectrum (e.g. ALMA, JWST, ATHENA) and in the gravitational waves domain (e.g. e-LISA, DECIGO).

The formation process of the first black holes is likely to produce a strong imprint on their observational signatures, as well as on their mass growth. 
The standard theory of Eddington-limited accretion predicts that black holes grow in mass over a time scale $\sim 0.045\epsilon_{0.1} \, \mathrm{Gyr}$, where $\epsilon_{0.1}$ is the matter-energy conversion factor normalized to the standard value of $10\%$. With black hole seeds of initial mass $\sim 100 \, \mathrm{\Msun}$, formed at the end of the very short ($\sim 1-10 \, \mathrm{Myr}$) lifetime of Pop III stars, it is, at best, challenging to explain recent observations of optically bright quasars with $M_{\bullet} \sim 10^{9-10} \, \mathrm{\Msun}$ at $z \sim 7$ \citep{Mortlock_2011, Wu_2015}.
An alternative, attractive solution is based on Massive Black Hole (MBH) seeds  ($10^{3-5} \, \mathrm{\Msun}$) appearing at $z\sim 10-15$, giving a jump start to the growth process \citep[e.g.,][]{2006ApJ...652..902S,Begelman_2006,Lodato_Natarajan_2006}.
Under specific conditions (\citealt{Bromm_Loeb_2003, Begelman_2006, Volonteri_2008, Shang_2010, Johnson_2012,Agarwal_2014}), the collapse of a primordial atomic-cooling halo may lead to the formation of MBHs with a birth mass function peaked at $M_{\bullet} \sim 2\times 10^5 \, \mathrm{\Msun}$ \citep{Ferrara_2014}. The subsequent gas accretion from the host halo leads to further growth into $\gsim 10^7 \, \mathrm{\Msun}$ objects.

The expected abundance of MBHs is largely unconstrained. \cite{Salvaterra_2012} and \cite{Treister_2013} provided upper limits of order $\rho_{\bullet} \lsim 10^{3-4} \, \mathrm{\Msun \, Mpc^{-3}}$, using the X-ray background and the stacked X-ray luminosity of high-redshift galaxies, respectively. Upper limits for the $z=6$ MBH mass density provided by \cite{Willott_2011}, \cite{Fiore_2012} and \cite{Cowie_2012} are even lower ($\rho_{\bullet} \lsim 10^{2-3} \, \mathrm{\Msun \, Mpc^{-3}}$), although none of these constraints take into account Compton-thick sources that can be buried deep inside dense nuclei in proto-galaxies. \cite{Yue_2013,Yue_2014} noted that if MBHs are responsible for the near-infrared background fluctuations, their high-redshift mass density should be comparable to the present-day value: $\rho_{\bullet}(z=0) \sim 2 \times 10^5 \, \mathrm{\Msun \, Mpc^{-3}}$, see \cite{Yu_2002}. The population of high-redshift MBHs would produce gravitational waves, during their collapse \citep{Pacucci_2015_GW} or ensuing MBH-MBH mergers \citep{GW3, Sesana2011}, detectable with upcoming observatories.

Our previous works \citep{Pacucci_2015,PVF} focused, through accurate 1D radiation-hydrodynamic simulations, on the dynamical evolution of $z \sim 10$ MBH seeds with initial mass $M_{\bullet} \sim 10^{3-6} \, \mathrm{\Msun}$, embedded in dark matter halos with total mass $M_h \sim 10^8 \, \mathrm{\Msun}$ and accreting in the standard Eddington-limited scenario, or including a model for super-Eddington accretion through slim disks (see also \citealt{Volonteri_2005, Volonteri_2014}).
In the present work we focus on their emission spectrum with three objectives: (i) predict the time evolution of the spectrum, (ii) assess the observability with current (Chandra Deep Field South, CDF-S) and future (JWST, ATHENA) surveys, and (iii) estimate the mass density $\rho_{\bullet}$ of high-redshift MBHs.

The outline of this paper is as follows. In $\S 2$ we describe the physical and numerical implementation, while in $\S 3$ we present our results for the spectral evolution of high-redshift MBHs. Finally, in $\S 4$ we provide some further discussion and the conclusions.
We adopt recent Planck \citep{Planck_2015} cosmological parameters throughout: $(\Omega_m, \Omega_{\Lambda}, \Omega_b, h, n_s, \sigma_8 )= (0.32, 0.68, 0.05, 0.67, 0.96, 0.83)$.

\section{Physical and Numerical Implementation}
\label{sec:physical_numerical_implementation}
The present work is based on radiation-hydrodynamic simulations post-processed with \texttt{CLOUDY}, a spectral synthesis code \citep{Cloudy}.

The physical framework is the following: a high-redshift ($z=10$) MBH seed with initial mass $10^5 \, \mathrm{\Msun}$ is located at the center of a dark matter halo with primordial composition and total mass $M_h \sim 10^8 \, \mathrm{\Msun}$ ($T_{\mathrm{vir}} \sim 10^4 \, \mathrm{K}$).
The MBH accretes mass from the inner parts (within $ \sim 10 \, \mathrm{pc}$) of the host halo until complete gas depletion. 

Our radiation-hydrodynamic code takes into account the frequency-integrated radiative transfer through the gas, with appropriate: (i) cooling and heating terms, (ii) matter-to-radiation coupling, and (iii) energy propagation through a two-stream approximation method. 
The code computes the accretion rate through the inner boundary of the simulation domain, from which we derive the total bolometric energy radiated by accretion, assuming a radiatively efficient or inefficient disk. The full {\em frequency-dependent} radiative transfer through the host halo is then performed in a post-processing step using \texttt{CLOUDY}. This code computes the detailed time-evolving spectrum emerging from the host halo using as input the matter distribution obtained from our radiation-hydrodynamics simulations and the realistic irradiation spectrum at the inner boundary, scaled to the appropriate bolometric luminosity. Additional details are given in sections 2.1 and 2.2.

\subsection{Dynamics and thermodynamics}
Our radiation-hydrodynamic code (see \citealt{Pacucci_2015} for an extensive description) solves the 1D spherically-symmetric equations of hydrodynamics and a frequency-integrated version of radiative transfer equations. The code evolves self-consistently the radial component of the standard system of ideal, non-relativistic Euler's equations (neglecting viscosity, thermal conduction and magnetic fields) for a gas accreting, with no angular momentum, onto the central MBH, assumed at rest and already formed at the time $t=0$, with a given initial mass $M_{\bullet}(t=0)$. The simulation domain spans from $0.1 \, \mathrm{pc}$ to $10 \, \mathrm{pc}$, largely encompassing the characteristic spatial scale for accretion, the Bondi radius:
\begin{equation}
R_B = \frac{GM_{\bullet}}{c_{s(\infty)}^2} \sim 3.0 \, \mathrm{pc} \, ,
\end{equation}
where $G$ is the gravitational constant and $c_{s(\infty)}=\sqrt{\gamma R T_{\infty}/\mu} \sim 12 \, \mathrm{km \, s^{-1}}$ is the sound speed at large distances from the accretion boundary; $\gamma = 5/3$ is the ratio of specific heats, $R$ is the gas constant, $T$ is the gas temperature and $\mu=1.15$ is the mean molecular weight for a primordial H-He composition gas with helium fraction $Y_P=0.24665$ \citep{Planck_2015} and no metals. 
For a $10^5 \, \mathrm{\Msun}$ object, the inner boundary of our spatial domain is $\sim 10^7$ times larger than the Schwarzschild radius and $\sim 10^5$ times larger than the centrifugal radius, i.e. the spatial scale below which deviations from spherical symmetry become important and an accretion disk may form.
Moreover, the angular momentum transfer in the outward direction of the accretion flow is very efficient, due to gravitational torques induced by dark matter and gas distributions of the halo \citep{Choi_2015}. The gas loses its angular momentum efficiently and flows well beyond its centrifugal barrier. Therefore, despite its simplifications, our 1D approach is significantly helpful in acquiring physical insights on the mechanisms regulating the black hole growth. For the same reason, neglecting viscosity, thermal conduction and magnetic fields is a safe choice, since they play an important role only on spatial scales comparable with the radius of the accretion disk.

The gas accretion through the inner boundary of our spatial domain produces an accretion rate $\dot{M}_\bullet$, which in turn generates an emitted luminosity $L$ via two different accretion models: (i) a standard Eddington-limited model in which $L=\epsilon c^2 \dot{M}$, $\epsilon=0.1$, and (ii) a radiatively inefficient model, the slim disk, in which $L\propto \ln(M_{\bullet})$ and $\epsilon \lsim 0.04$ is a function of $\dot{M}_{\bullet}$.
The main physical quantity which determines the properties of the accretion disk, and consequently the radiative efficiency, is the accretion rate. 
Accretion of gas at moderate rates ($0.01 \lsim f_{Edd} \lsim 1$) is expected to form a radiatively efficient, geometrically thin and optically thick accretion disk, which is typically modelled with the standard $\alpha$-disk model \citep{Shakura_Sunyaev_1973}. In a Shakura \& Sunyaev disk the radiative efficiency is determined only by the location of the innermost stable circular orbit, which in turn depends only on the spin of the black hole, and it varies between $\sim 6 \%$ and $\sim 32 \%$ for a non-spinning and a maximally spinning black hole, respectively \citep{Thorne_1974}. 
In a super-critical ($f_{Edd} > 1$) accretion environment the structure of the accretion disk is, instead, expected to be geometrically and optically thick and radiatively inefficient (but see \citealt{Jiang_2014, McKinney_2015}).
The most common solution proposed for such accretion flows is the slim disk (\citealt{Paczynski_1982, Abramowicz_1988, Mineshige_2000, Sadowski_2009, Sadowski_2011, McKinney_2014}). In these scenarios, part of the energy produced inside the disk is advected inwards (see, e.g., \citealt{Abramowicz_2013, Lasota_2015}) out to a spatial scale named the photon trapping radius ($R_{tr} \sim R_s f_{Edd}$, where $R_s$ is the Schwarzschild radius). Therefore, only a fraction of the photons produced in the accretion disk is able to free stream out of $R_{tr}$: consequently, the effective radiation and radiation pressure escaping to infinity is decreased (see e.g. \citealt{Begelman_1978, Ohsuga_2002}).
While the slim disk solution is the simplest and most tested model for super-critical accretion, alternatives exist, e.g the ZEro-BeRnoulli Accretion (ZEBRA, \citealt{Coughlin_2014}) and the ADiabatic Inflow-Outflow Solutions (ADIOS, \citealt{Blandford_1999, Begelman_2012}) models. These theoretical models also include a parameter which describes the fraction of the inflowing mass which is lost due to radiation pressure.

Radiation pressure accelerates the gas via:
\begin{equation}
a_{rad}(r) = \frac{\kappa(\rho, T) L(r)}{4 \pi r^2 c} \, ,
\end{equation}
where the gas opacity $\kappa(\rho, T)$ includes Thomson and bound-free terms, with the inclusion of a temperature dependence \citep{Begelman_2008}.
The radiation pressure may be able to temporarily interrupt the gas inflow, resulting in an intermittent accretion and outflows. The physical parameters regulating this occurrence are investigated in \cite{PVF}.

We assume that the gas initially follows the isothermal ($T \sim 10^4 \, \mathrm{K}$) density profile derived from the simulations in \cite{Latif_2013}, approximated by the functional form:
\begin{equation}
\rho(r) = \frac{\rho_0}{1+(r/a)^2} \, ,
\end{equation}
where $a$ is the core radius and $\rho_0$ is the central density. To understand how the matter distribution influences both the accretion and the emerging spectrum (through the hydrogen column density), we implemented two different density profiles, both of them yielding a gas mass $\sim 10^7 \, \mathrm{\Msun}$: (i) a high density profile (HDP) with a central density $\rho_0 = 10^{-12} \mathrm{g \, cm^{-3}}$ and a core radius $a=0.002 \, \mathrm{pc}$, and (ii) a low density profile (LDP) with a central density $\rho_0 = 10^{-18} \mathrm{g \, cm^{-3}}$ and a core radius $a=2 \, \mathrm{pc}$. The LDP may be thought as the density profile resulting after the formation of a MBH of mass $\sim 10^5 \, \mathrm{\Msun}$ at the halo center (see \citealt{Latif_2013c, Latif_2014b}), while in the HDP case the seed formed is very small ($\lsim 10^3 \, \mathrm{\Msun}$). A density floor of $10^{-24} \, \mathrm{g \, cm^{-3}} \sim 1 \, \mathrm{cm^{-3}}$, a factor at least $10^6$ times smaller than the central density for both profiles, is imposed for numerical stability reasons. In summary, we have four models: standard accretion - LDP, standard accretion - HDP, slim disk accretion - LDP, slim disk accretion - HDP.

\subsection{Spectrum}
Our code evolves the system in time until gas depletion and provides \texttt{CLOUDY} with data to compute the spectrum emerging from the host halo.
This code takes into account: (i) the spatial profiles for hydrogen number density $n_H(r)$ and temperature $T(r)$, (ii) the source spectrum of the central object, and (iii) the bolometric luminosity $L$ of the source, computed self-consistently from $\dot{M}_{\bullet}$.
The spherical cloud of gas is assumed to be metal-free, so that only H and He recombination lines are present. The addition of a small amount of metals, formed by the first Pop III stars, would increase the number of lines and the absorption of high-energy photons.
The source spectral energy distribution is taken from \cite{Yue_2013} and can be described as the sum of three components: (i) a multi-colour blackbody, (ii) a power-law, and (iii) a reflection component.
The source spectrum is extended from far-infrared to hard X-ray ($\sim 1 \, \mathrm{MeV}$).

\section{Results}
\label{sec:results}
We describe in the following the standard accretion - LDP case, and discuss the differences with the other cases when needed. 

\begin{figure}
\vspace{-1\baselineskip}
\hspace{-0.5cm}
\begin{center}
\includegraphics[angle=0,width=0.50\textwidth]{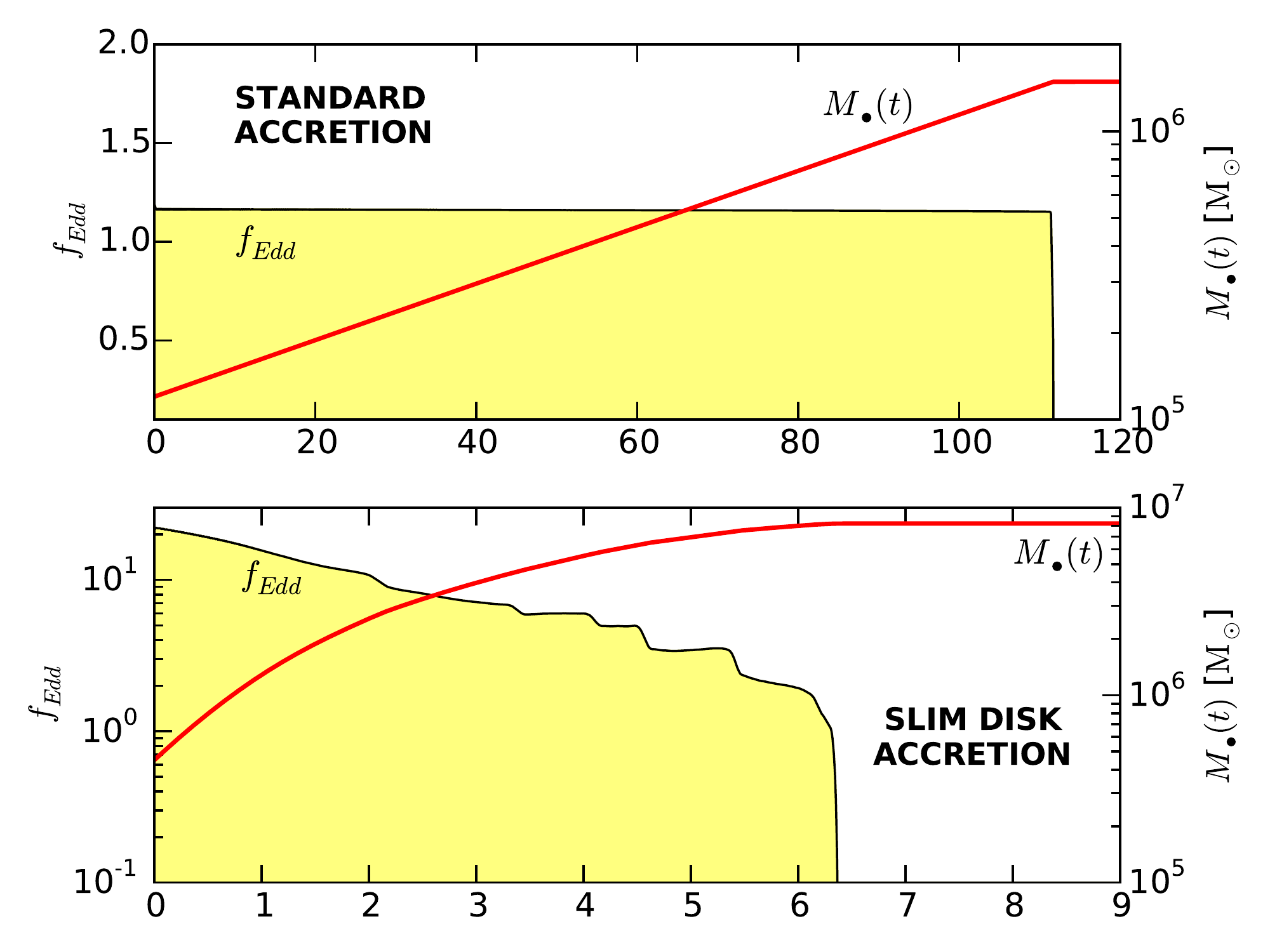}
\caption{Time evolution of the Eddington factor $f_{Edd}$ and of the black hole mass $M_{\bullet}$, in the standard accretion - LDP (top) and the slim disk accretion - LDP (bottom) scenarios. The Eddington ratio is reported as a running average over periods of $\sim 0.1 \, \mathrm{Myr}$.}
\label{fig:f_Edd_M}
\end{center}
\end{figure}

The balance between the inward gravitational acceleration and the outward radiation pressure keeps the accretion rate close to the Eddington level ($f_{Edd} \approx 1.2$) for most of the time, ensuring a continuous accretion. The top panel of Fig. \ref{fig:f_Edd_M} shows the time evolution of $f_{Edd}$ and of $M_{\bullet}$ in this accretion scenario. Defining the duty cycle, ${\cal D}$, as the fraction of time spent accreting during the total evolutionary time of the system, we find that in the LDP cases ${\cal D} =1$, while other accretion scenarios may be characterized by quiescent phases (${\cal D} < 1$, see Table \ref{tab:outline}). The system evolves for $\sim 120 \, \mathrm{Myr}$, until complete depletion of the $\sim 10^7 \, \mathrm{\Msun}$ gas reservoir within $\sim 10 \, \mathrm{pc}$. However, this time scale is a lower limit, since in a real galaxy the gas would extend much further.

\begin{figure}
\vspace{-1\baselineskip}
\hspace{-0.5cm}
\begin{center}
\includegraphics[angle=0,width=0.50\textwidth]{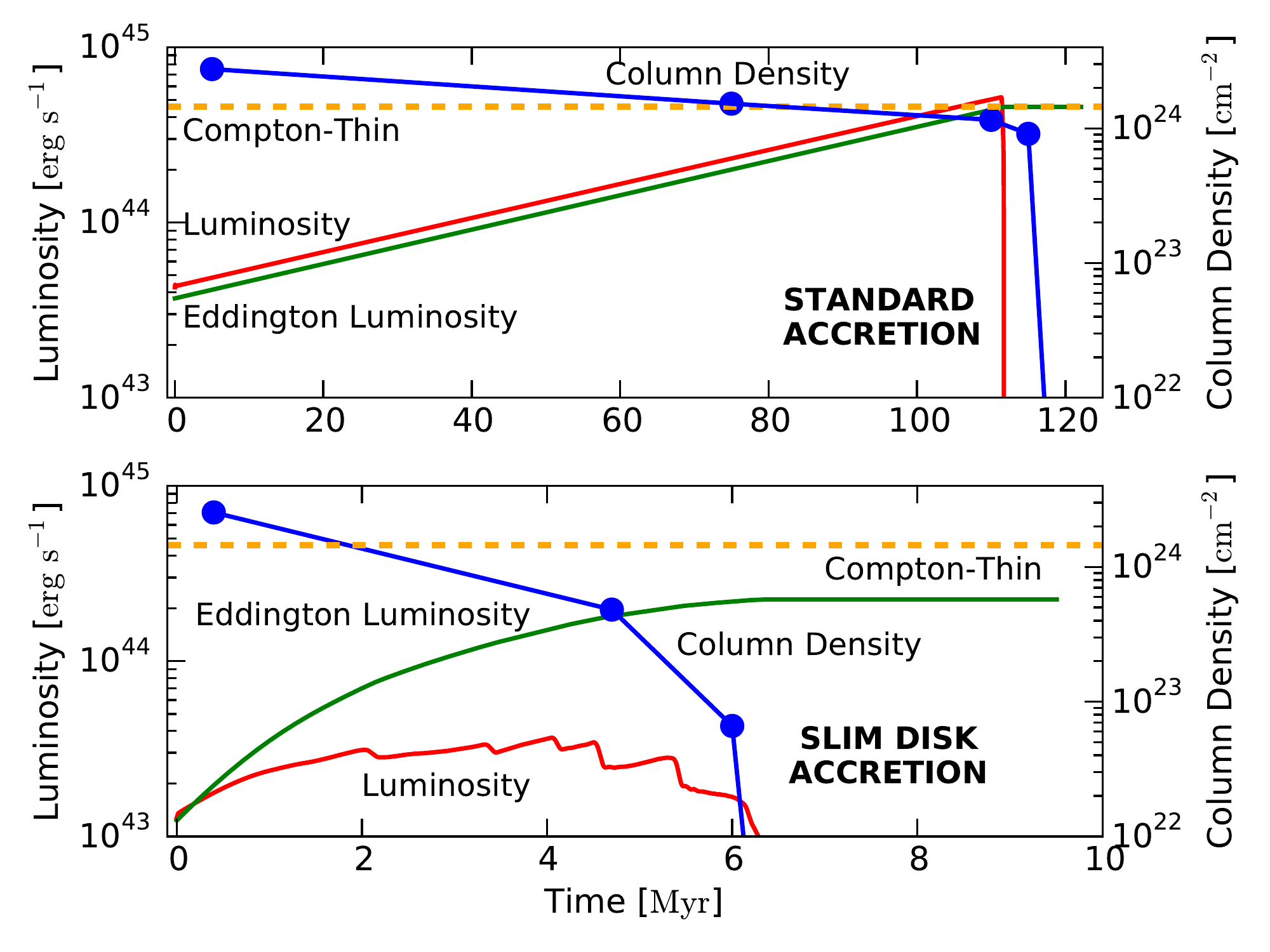}
\caption{Time evolution of the bolometric luminosity emitted at the inner boundary, before traversing the host halo gas, in the standard accretion - LDP (top) and the slim disk accretion - LDP (bottom) scenarios. The luminosity is reported as a running average over periods of $\sim 0.1 \, \mathrm{Myr}$. The corresponding Eddington luminosity, $L_{Edd} \propto M_{\bullet}$, and the values of the hydrogen column density (blue curves, right axis) are also shown.}
\label{fig:luminosity_cd}
\end{center}
\end{figure}

Fig. \ref{fig:luminosity_cd} (top panel) shows the time evolution of the bolometric luminosity emitted at the inner boundary, before traversing the host halo gas.
The corresponding Eddington luminosity:
\begin{equation}
L_{Edd} \equiv \frac{4 \pi G M_{\bullet} c}{\kappa_T} \propto M_{\bullet} \, ,
\end{equation}
with $\kappa_T$ being the Thomson opacity, is also shown for comparison.
The luminosity increases for the first $\sim 115 \, \mathrm{Myr}$, reaching a peak of $\sim 5\times 10^{44} \, \mathrm{erg \, s^{-1}}$ and is, on average, mildly super-Eddington as long as the amount of gas is sufficient to sustain this accretion rate. Afterwards, the luminosity plummets, when all the available gas has been consumed.
For comparison, the interested reader is referred to \cite{Pacucci_2015}, where a plot of the same physical quantities is shown in Fig. 8.
Moreover, Fig. \ref{fig:luminosity_cd} also shows the values of the hydrogen column density, $N_H$, computed at selected times ($t_s=5, \, 75, \, 110, \, 115, \, 120 \, \mathrm{Myr}$) when also the emergent spectrum is computed. The system is initially Compton-thick ($N_H \gsim 1.5 \times 10^{24} \, \mathrm{cm^{-2}}$, see the horizontal line). As the gas is progressively accreted by the MBH, $N_H$ steadily decreases before a sudden drop ($t_s \approx 115$ Myr) corresponding to the remaining gas evacuation by radiation pressure. Table \ref{tab:outline} lists duty cycles and accretion time scales for all four models.

The bottom panels of Fig. \ref{fig:f_Edd_M} and Fig. \ref{fig:luminosity_cd} show the time evolution of $f_{Edd}$, $M_{\bullet}$, $L$ and $N_H$ in the slim disk - LDP case. In this accretion scenario, the Eddington rate reaches high values ($f_{Edd} \sim 20$) only for a short amount of time ($\sim 2 \, \mathrm{Myr}$), while afterwards the accretion rates are sub-Eddington. The evolution is much more rapid (the available gas is consumed in $\sim 6 \, \mathrm{Myr}$, a factor of $\sim 16$ faster than in the standard case) and the host halo becomes Compton-thin ($N_H \lsim 1.5 \times 10^{24} \, \mathrm{cm^{-2}}$) in $\lsim 2 \, \mathrm{Myr}$.
As already noted in \cite{PVF}, in the slim disk accretion - LDP case, the black hole is able to accrete up to $\sim 80 \%$ of the gas mass within $\sim 10 \,  \mathrm{pc}$, with respect to the $\sim 15 \%$ in the standard accretion scenario, and the black hole grows to $\sim 8 \times 10^6 \, \mathrm{\Msun}$ in $\sim 6 \, \mathrm{Myr}$, while in the standard case the black hole grows to $\sim 1.5 \times 10^6 \, \mathrm{\Msun}$ in $\sim 100 \, \mathrm{Myr}$. 

\subsection{Spectral evolution}
\begin{figure}
\vspace{-1\baselineskip}
\hspace{-0.5cm}
\begin{center}
\includegraphics[angle=0,width=0.50\textwidth]{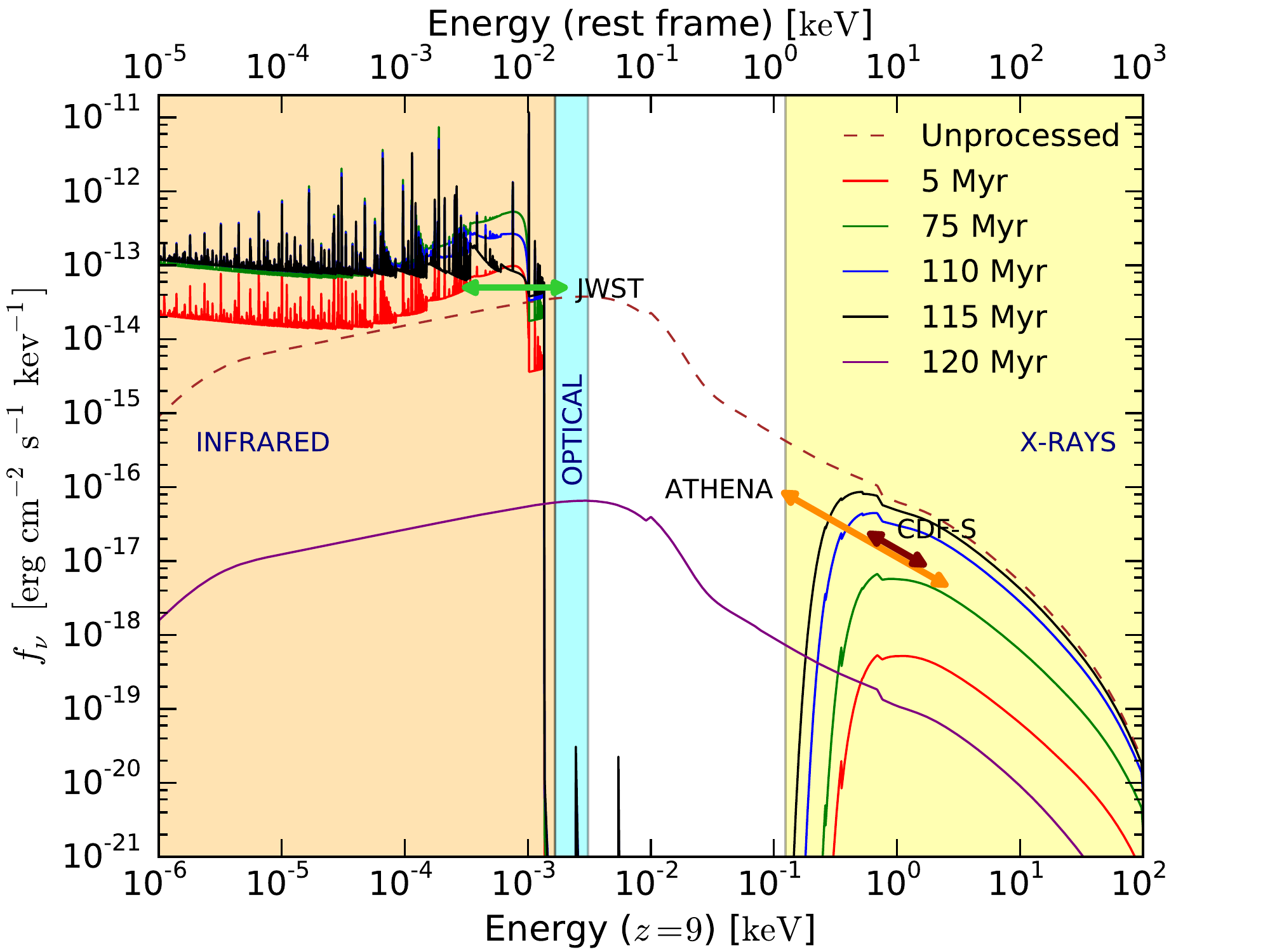}
\caption{Time evolution of the spectrum emerging from the host halo for a source located at $z=9$, in the standard accretion - LDP case. The infrared, optical and X-ray bands are highlighted with shaded regions, while the unprocessed spectrum is reported, at peak luminosity ($t=115 \, \mathrm{Myr}$), with a dashed line. The flux limits for future (JWST, ATHENA) and current (CDF-S) surveys are also shown.}
\label{fig:spectrum}
\end{center}
\end{figure}

The time-evolving spectrum emerging from the host halo, in the standard accretion - LDP case, is shown in Fig. \ref{fig:spectrum} at $z=9$, i.e. $\sim 100 \, \mathrm{Myr}$ after the beginning of the simulation ($z=10$).
The spectrum is composed by: (i) the continuum emitted by the source and attenuated by the gas, and (ii) the diffuse emission of the gas. Most of the energy emerges in the \emph{observed } infrared and X-ray bands. The latter is characterized by a bell-shaped spectrum peaked around $1 \, \mathrm{keV}$, while in the infrared band a large number of H-He nebular lines is present.
Photons with frequency shortwards than the Ly$\alpha$ line are absorbed by the intervening matter at column densities $N_H \gsim 10^{23} \, \mathrm{cm^{-2}}$ and reprocessed at lower energies, boosting the infrared emission of the halo. 
X-ray emission occurs predominantly within the rest-frame energy range $4 \, \mathrm{keV} < E_{\gamma} < 10 \, \mathrm{keV}$. The mean free path of such photons is much larger than the Hubble radius at $z=9$: $\lambda_X \gsim 8 \, \mathrm{Gpc} \gg R_H(z=9) = 165 \, \mathrm{Mpc}$. Hence growing MBH seeds negligibly contribute to reionization. 
The increase with time of the continuum normalization in the X-ray is mainly due to the progressive rise of the bolometric luminosity of the central object.
The ratio between the infrared and the X-ray continua depends on the column density, since in the Compton-thick case ($N_H \gsim 1.5 \times 10^{24} \, \mathrm{cm^{-2}}$) the high-energy frequencies are heavily absorbed and reprocessed at lower energies, leading to an overall increase of the infrared emission.
When the gas becomes Compton-thin at $\sim 75 \, \mathrm{Myr}$, the X-ray continuum progressively increases, while the infrared one starts to decrease.
Approaching the complete gas depletion within $\sim 10 \, \mathrm{pc}$, at $\sim 120 \, \mathrm{Myr}$, the column density is so low (see Fig. \ref{fig:luminosity_cd}, top panel) that the outgoing radiation is nearly unimpeded (the emerging spectrum is very similar to the source spectrum, reported, at peak luminosity, $t=115 \, \mathrm{Myr}$, as a dashed line) and the continuum normalization drops by $\sim 3$ ($\sim 4$) orders of magnitude in the X-ray (infrared) band.
\subsection{High-redshift MBH mass density}
The approximate flux thresholds for two future-generation observatories (JWST\footnote{For a NIRcam observation with a Signal-to-Noise ratio of $10$ and a total integration time of $10^4 \, \mathrm{s}$.} in the infrared band, ATHENA\footnote{For a $3\sigma$ detection with a total integration time of $3 \times 10^5 \, \mathrm{s}$.} in the X-ray) and for the CDF-S\footnote{Ultra-deep survey in the X-ray with a total integration time of $4 \times 10^6 \, \mathrm{s}$.} survey are shown in Fig. \ref{fig:spectrum}. 
We predict that the JWST will be able to observe most ($\gsim 95\%$) of the accretion process onto a $10^5 \, \mathrm{\Msun}$ MBH seed up to a comoving distance corresponding to $z \sim 25$, while ATHENA will only detect $\sim 25\%$ of the total evolution, around the peak luminosity, up to $z \sim 15$.

Comparing the predicted peak flux (in the $1 \, \mathrm{keV}$ observed band) with the CDF-S sensitivity, we find that this ultra-deep survey could have observed the accretion process onto a typical MBH up to $z_{max} \sim 15$.
In the CDF-S survey, $N_{C} = 3$ AGN candidates at $z \gsim 6 \equiv z_{min}$ have been identified (\citealt{Giallongo_2015}, but see also \citealt{Weigel_2015} where the authors question some of these candidates) inside a sky region of $\sim 170 \, \mathrm{arcmin^2}$.
An upper limit for the number density of MBH seeds can be derived as:
\begin{equation}
n_{\bullet} (6<z<15) \lsim \frac{N_C}{\Omega_{CDF}{\cal V}{\cal D}{\cal F}} \, ,
\end{equation}
where $\Omega_{CDF} = 1.1 \times 10^{-6}$ is the sky fraction observed by the CDF-S, ${\cal V}=2.3 \times 10^{12} \, \mathrm{Mpc^3}$ is the comoving volume of the Universe between $z_{min}$ and $z_{max}$, ${\cal D}$ is the duty cycle for accretion and ${\cal F} \equiv t_{det}/t_{obs}$ is the fraction of time during which the object is detectable within the time frame $t_{obs}(6<z<15)=670 \, \mathrm{Myr}$, assuming a single episode of MBH growth at these redshifts\footnote{This calculation assumes spherical symmetry of the host halo and an isotropic irradiation of the MBH. If the density along the poles is much lower, then a fraction of the sources would be completely unobscured.}.
The predicted flux, within the spectral range $0.5 - 2.0 \, \mathrm{keV}$, is above the CDF-S sensitivity between $\sim 85 \, \mathrm{Myr}$ and $\sim 115 \, \mathrm{Myr}$, so that $t_{det}= 30 \, \mathrm{Myr}$ and ${\cal F} =0.045$. Using these values we obtain the following upper limit for the number density of MBH seeds:
\begin{equation}
n_{\bullet} (6<z<15) \lsim \frac{2.5 \times 10^{-5}}{{\cal D}} \left( \frac{0.045}{{\cal F}} \right) \, \mathrm{Mpc^{-3}} \, .
\end{equation}
Considering MBHs of initial mass $10^5 \, \mathrm{\Msun}$ growing up to $\sim 10^7 \, \mathrm{\Msun}$ \citep{Pacucci_2015}, we finally obtain the following upper limit for the MBH seeds mass density: 
\begin{equation}
\rho_{\bullet} (6<z<15)  \lsim \frac{2.5 \times 10^{2}}{{\cal D}} \left( \frac{0.045}{{\cal F}} \right) \, \mathrm{\Msun \, Mpc^{-3}} \, .
\end{equation}

The discussion so far has been limited to the scenario in which a MBH with an initial mass close to the peak of the birth mass function devised in \cite{Ferrara_2014} accretes gas from a LDP host halo in the Eddington-limited mode. However, this is not the only possible scenario.
Table \ref{tab:outline} provides a general outline of the accretion history and CDF-S observability for a MBH seed with initial mass $10^{5} \, \mathrm{\Msun}$ in three additional scenarios: standard accretion - HDP, slim disk accretion - LDP and slim disk accretion - HDP. 
Table \ref{tab:outline} includes the depletion time $t_{end}$, the duty cycle ${\cal D}$, the detection time $t_{det}$ with its related value of ${\cal F}$ and the upper limit on $\rho_{\bullet}$.
In the Eddington-limited cases the depletion times are lower limits since, while our simulations are run in isolation, in a cosmological framework the halo growth by mergers and accretion would not be negligible within $\sim 100 \, \mathrm{Myr}$. In the slim disk case, given the depletion times of order $\sim 10 \, \mathrm{Myr}$, the hierarchical growth of the halo plays a minor role.
\begin{table*}
\begin{minipage}{170mm}
\begin{center}
\caption{Accretion history and CDF-S observability for a MBH seed with initial mass $M_{\bullet} = 10^{5} \, \mathrm{\Msun}$ in the four indicated accretion scenarios.}
\label{tab:outline}
\begin{tabular}{|c|c|c|c|c|c|c|}
\hline\hline
Accretion scenario & Observable & $t_{end} \, \mathrm{[Myr]}$ & ${\cal D}$ & $t_{det} \, \mathrm{[Myr]}$ & ${\cal F}$ & $\rho_{\bullet} \, [\mathrm{\Msun \, Mpc^{-3}}]$\\
\hline
Standard accretion - LDP  & YES & $ 120 $    &$1.0$  & $30$ & $0.045$ & $\lsim 2.5 \times 10^{2}$ \\
Standard accretion - HDP & YES & $240 $    & $0.4$ & $110$ & $0.16$  & $\lsim  1.8 \times 10^{2}$  \\
Slim disk accretion - LDP & NO &$7$    & $1.0$ & $0$ & $0$ & No constraints\\
Slim disk accretion - HDP &YES & $12$    & $1.0$ & $1$ & $0.0015$  & $\lsim  7.6 \times 10^{3}$\\
\end{tabular}
\end{center}
\end{minipage}
\end{table*} 

\begin{figure}
\vspace{-1\baselineskip}
\hspace{-0.5cm}
\begin{center}
\includegraphics[angle=0,width=0.50\textwidth]{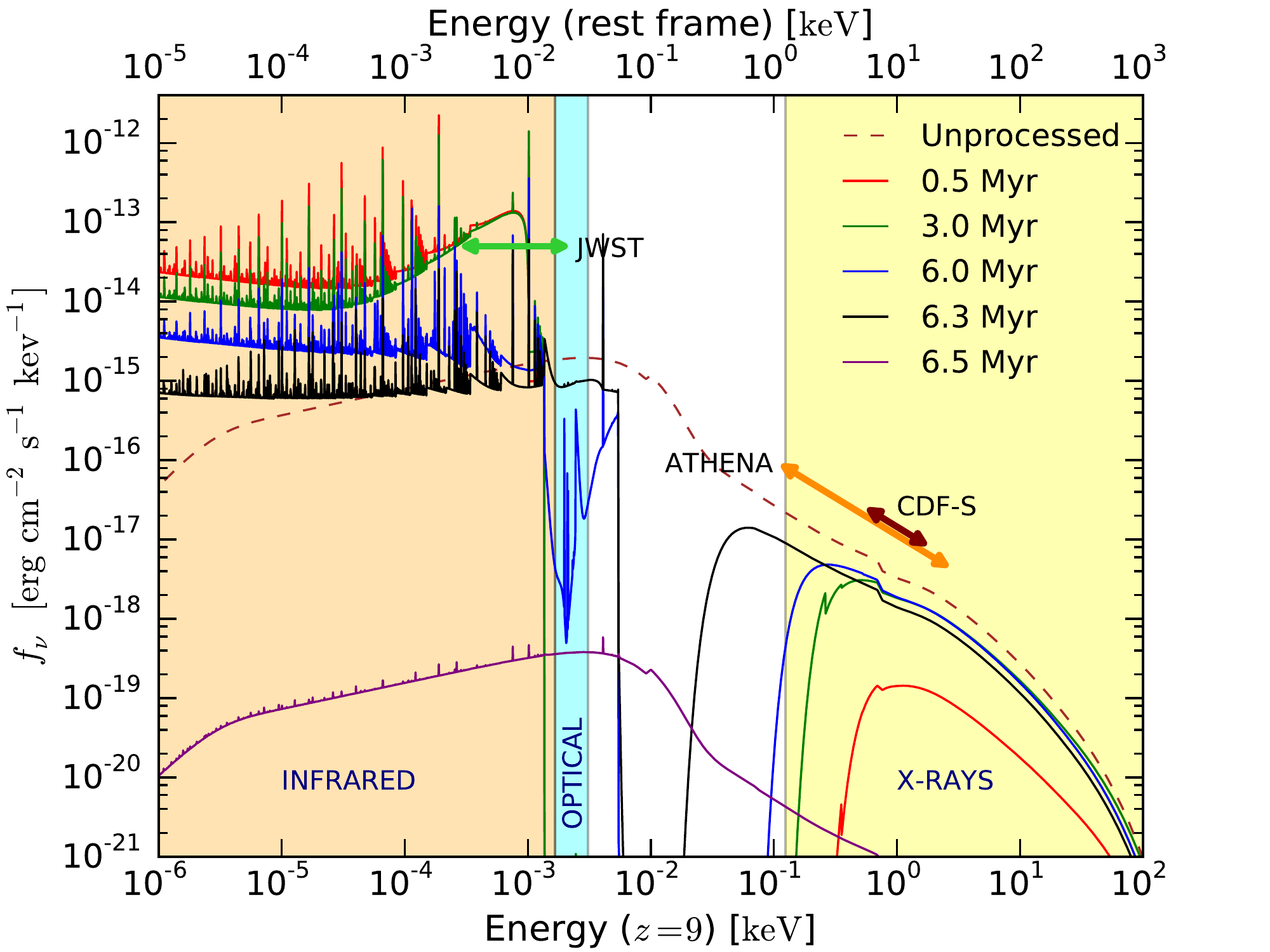}
\caption{As in Fig. \ref{fig:spectrum}, but for the slim disk accretion - LDP case.}
\label{fig:spectrum_SD}
\end{center}
\end{figure}

Interestingly, the accretion process is undetectable by the CDF-S in the slim disk - LDP case, whose time-evolving spectrum is shown in Fig \ref{fig:spectrum_SD}. Therefore no constraints on $\rho_{\bullet}$ can be obtained in this case. 
The slim disk - HDP case is instead observable, albeit only for a very short time, $t_{det} \sim 1 \, \mathrm{Myr}$, due to the larger accretion rates in this scenario, which produce a sufficiently high luminosity despite radiation trapping. The very low value of ${\cal F} \sim 1.5 \times 10^{-3}$ produces an upper limit for $\rho_{\bullet}$ higher than in other accretion scenarios by a factor $\sim 35$.

Let us now focus on the reason why the slim disk case leads to a very different value for $\rho_{\bullet}$ with respect to the standard accretion case.
The amount of mass which is available for accretion is equal in both cases ($\sim 10^7 \, \mathrm{\Msun}$) and, to a first-order approximation, also the mass actually accreted is similar. This is not strictly true, since the radiation pressure is more efficient in the standard case in creating mass outflows \citep{PVF}, but the produced outflows are fairly weak regardless of the accretion scenario.
The real difference between the slim disk and the standard cases is due to radiation trapping, which decreases the effective bolometric luminosity (i.e. the luminosity escaping to infinity) in the former case, with respect to the latter.
Since the accreting black hole is intrinsically fainter in the slim disk case, it will be observable for a smaller fraction of time, ${\cal F}$ (with the extreme ${\cal F}=0$ for the LDP case), during its evolution: this eventually leads to a larger upper limit on $\rho_{\bullet}$.

The initial mass of the MBH seed influences the time scale of the process and its observability. If, for instance, the high-redshift population of MBHs is characterized by a larger average mass, the evolutionary time scale would be smaller (since $t_{end} \propto \dot{M}_{Edd}^{-1} \propto M_{\bullet}^{-1}$, where $\dot{M}_{Edd}$ is the Eddington accretion rate), but the emitted luminosity would be higher (since $L \propto L_{Edd} \propto M_{\bullet}$).
The effect on ${\cal F}$ may vary from case to case, but in general a larger average mass is likely to decrease ${\cal F} \equiv t_{det}/t_{obs}$ since $t_{det} \leq t_{end} \propto M_{\bullet}^{-1}$. This would lead to a less stringent upper limit on $\rho_{\bullet}$.

\section{Discussion and Conclusions}
\label{sec:disc_concl}
Using a combination of radiation-hydrodynamic and spectral synthesis codes, we have investigated the time-evolving spectral energy distribution of an accreting $z \sim 10$ MBH. The MBH seed, whose initial mass is $10^5 \, \mathrm{\Msun}$, is embedded in a dark matter halo of total mass $10^{8} \, \mathrm{\Msun}$. Employing two gas density profiles and two accretion modes (Eddington-limited and slim disk) we simulated the system until complete gas depletion and we accurately calculated the time-evolving spectrum of the radiation emerging from the host halo.
The main results of this work are summarized in the following.
\vspace{-0.2cm}
\begin{itemize}
\item The spectrum of the emerging radiation, for a MBH observed at $z=9$, is dominated by the infrared-submm ($1-1000 \, \mathrm{\mu m}$) and X-ray ($0.1 - 100 \, \mathrm{keV}$) bands.  Photons with frequency shortwards than the Ly$\alpha$ line are absorbed by the intervening matter at column densities $N_H \gsim 10^{23} \, \mathrm{cm^{-2}}$ and reprocessed at lower energies, in the infrared band.  Due to the very large mean free path $\lambda_X > 1 \, \mathrm{Gpc}$ of X-ray photons, growing MBH seeds negligibly contribute to reionization. The continuum normalization is set by: (i) the bolometric luminosity of the source and (ii) the column density of the host halo. The former determines the overall normalization, while the latter determines the ratio between the low-energy and the high-energy continua.
\item Our predictions show that the JWST will detect in the infrared a fraction $\gsim 95\%$ of the accretion process onto a typical MBH seed observed at $z \sim 9$, while ATHENA should observe it in the high-energy bands only around the peak luminosity, a fraction $\sim 25\%$ of the total evolution. Similarly, long-exposure surveys in the X-ray, like the CDF-S, could have already observed the accretion process on a $z \sim 9$ object for a comparable fraction of time.
The redshift of the sources sets their luminosity distance, hence influences their detectability. For instance, the standard accretion - LDP system is observable by the CDF-S for $\sim 41\%$ of the time at $z_{min}=6$, while it becomes undetectable at $z \gsim z_{max}=15$.
\item From the $z \gsim 6$ candidates detected in the CDF-S survey \citep{Giallongo_2015} we estimate the following upper limits on the $z \gsim 6$ MBH mass density: (a) $\rho_{\bullet} \lsim 2.5 \times 10^{2} \, \mathrm{\Msun \, Mpc^{-3}}$ assuming Eddington-limited accretion; (b) $\rho_{\bullet} \lsim 7.6 \times 10^{3} \, \mathrm{\Msun \, Mpc^{-3}}$ if accretion occurs in the slim disk, highly obscured mode. However, the accretion process is undetectable with the CDF-S sensitivity in the slim disk accretion - LDP case, due to the flux suppression caused by radiation trapping, and no constraints on  $\rho_{\bullet}$ can be given.
\end{itemize}

Very recently, we proposed that the first detection of a high-redshift MBH seed could have already occurred. Indeed, in \cite{Pallottini_Pacucci_2015} we showed that the observational features of CR7 \citep{Sobral_2015}, a bright Ly$\alpha$ emitter at $z=6.604$, may be explained by accretion onto a MBH of initial mass $\sim 10^5 \, \mathrm{\Msun}$.

For  Eddington-limited accretion, our upper limit, $\rho_{\bullet} \lsim 2.5 \times 10^{2} \, \mathrm{\Msun \, Mpc^{-3}}$, is compatible with the one set by \cite{Cowie_2012} using observations of faint X-ray sources in the CDF-S, while it is more stringent than limits by \cite{Willott_2011}, \cite{Fiore_2012} and \cite{Treister_2013} ($\rho_{\bullet} \lsim 10^{3} \, \mathrm{\Msun \, Mpc^{-3}}$) and particularly by \cite{Salvaterra_2012} ($\rho_{\bullet} \lsim 10^{4} \, \mathrm{\Msun \, Mpc^{-3}}$, using the unresolved X-ray emission). The current observational constraints, however, do not take into account heavily buried, Compton-thick objects or radiatively inefficient accretion. Recently, indeed, \cite{Comastri_2015} suggested that recent revisions to the local SMBHs mass density, up to $\rho_{\bullet}(z=0) \sim 10^6 \, \mathrm{\Msun \, Mpc^{-3}}$, seem to imply that a significant fraction of the local SMBHs have grown in heavily buried, Compton-thick phases, or by radiatively inefficient accretion. 
Our model is a first step towards testing the role of Compton-thick or radiatively inefficient phases in the early growth of MBHs.

In the present paper, we provided a general picture of the interconnection between the main accretion mode at work in the high-redshift Universe and the black hole mass density. To summarize our rationale: assuming that, for MBHs at $z\gsim 6$, the main accretion channel is the standard, Eddington-limited one, we are able to provide an upper limit on $\rho_{\bullet}$ which is consistent (and competitive) with current estimates. In the standard disk scenario, MBHs have grown in heavily buried, Compton-thick phases for about $70\%$ of the total evolution time (e.g., $\sim 70 \, \mathrm{Myr}$ for the LDP), but they are long-lived and intrinsically bright, and visible for a substantial amount of time. Assuming instead that at super-critical accretion rates the accretion disk thickens, with radiation trapping playing a significant role, the suppressed radiative efficiency leads to a much lower intrinsic luminosity. These sources are intrinsically faint, not obscured: the Compton-thick phase is short ($\sim 2 \, \mathrm{Myr}$, $30\%$ of the total evolution time for the LDP) because the obscuring gas is consumed rapidly. Our slim disk simulations suggest, as one would expect, that these short-lived and fainter MBHs are more difficult to detect in current surveys compared to brighter objects accreting in the Eddington-limited mode. As a consequence, the upper limit on $\rho_{\bullet}$ is inevitably higher than currently predicted, up to a factor $\sim 35$. 

\vspace{0.5 cm}
\noindent
MV acknowledges support from a Marie Curie FP7-Reintegration-Grant (PCIG10-GA-2011-303609).
GD acknowledges support from the Centre Nationale d'Etudes Spatiales (CNES).

\section{Supplementary Figures}
In the following page we include additional figures showing the time evolution of the emerging spectrum (in $\nu f_{\nu}$ units) for the four accretion scenarios discussed so far.
\clearpage

\begin{figure}
\vspace{-1\baselineskip}
\hspace{-0.5cm}
\begin{center}
\includegraphics[angle=0,width=0.50\textwidth]{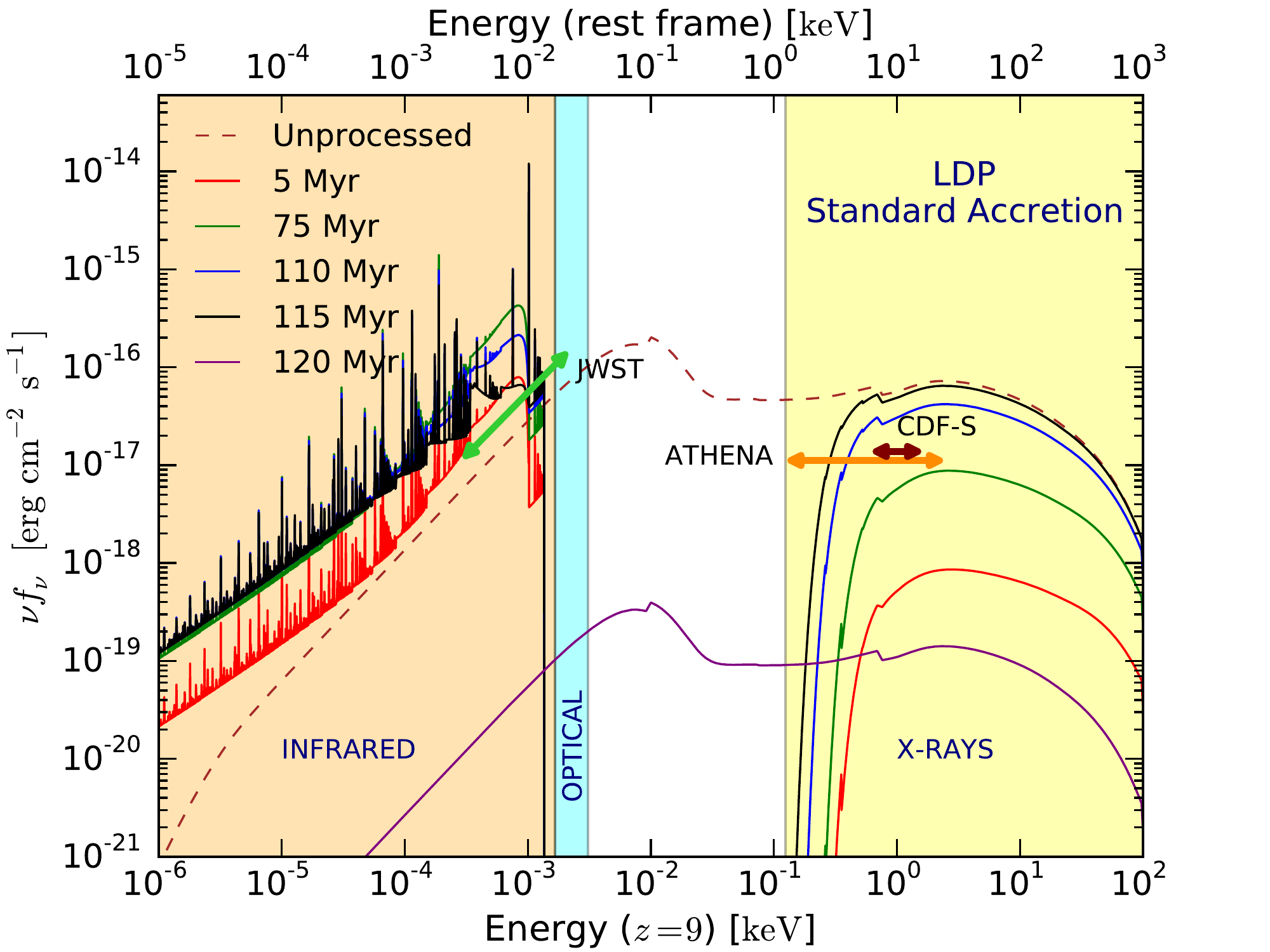}
\caption{Standard accretion - LDP case.}
\end{center}
\end{figure}

\begin{figure}
\vspace{-1\baselineskip}
\hspace{-0.5cm}
\begin{center}
\includegraphics[angle=0,width=0.50\textwidth]{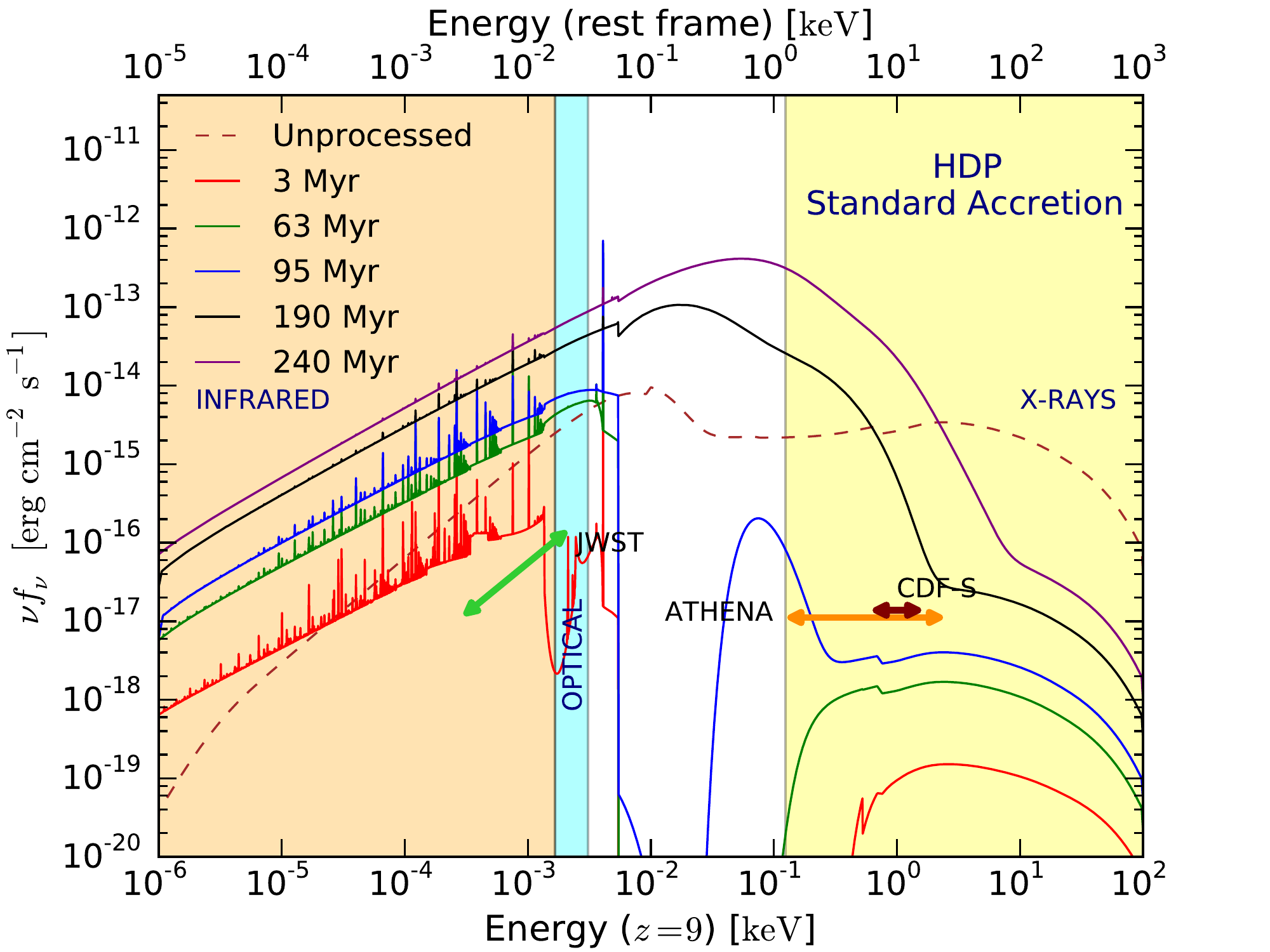}
\caption{Standard accretion - HDP case.}
\end{center}
\end{figure}

\begin{figure}
\vspace{-1\baselineskip}
\hspace{-0.5cm}
\begin{center}
\includegraphics[angle=0,width=0.50\textwidth]{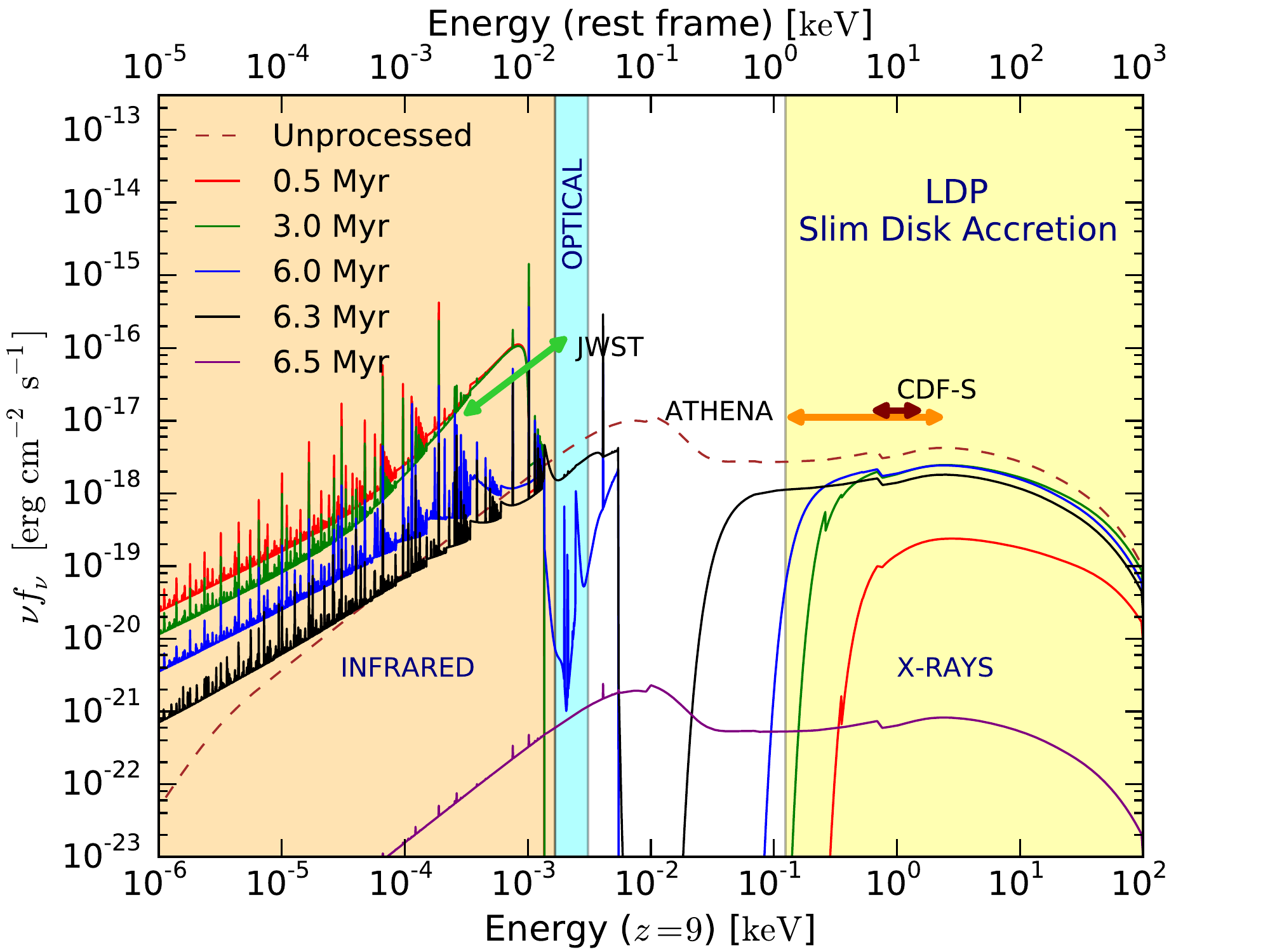}
\caption{Slim disk accretion - LDP case.}
\end{center}
\end{figure}

\begin{figure}
\vspace{-1\baselineskip}
\hspace{-0.5cm}
\begin{center}
\includegraphics[angle=0,width=0.50\textwidth]{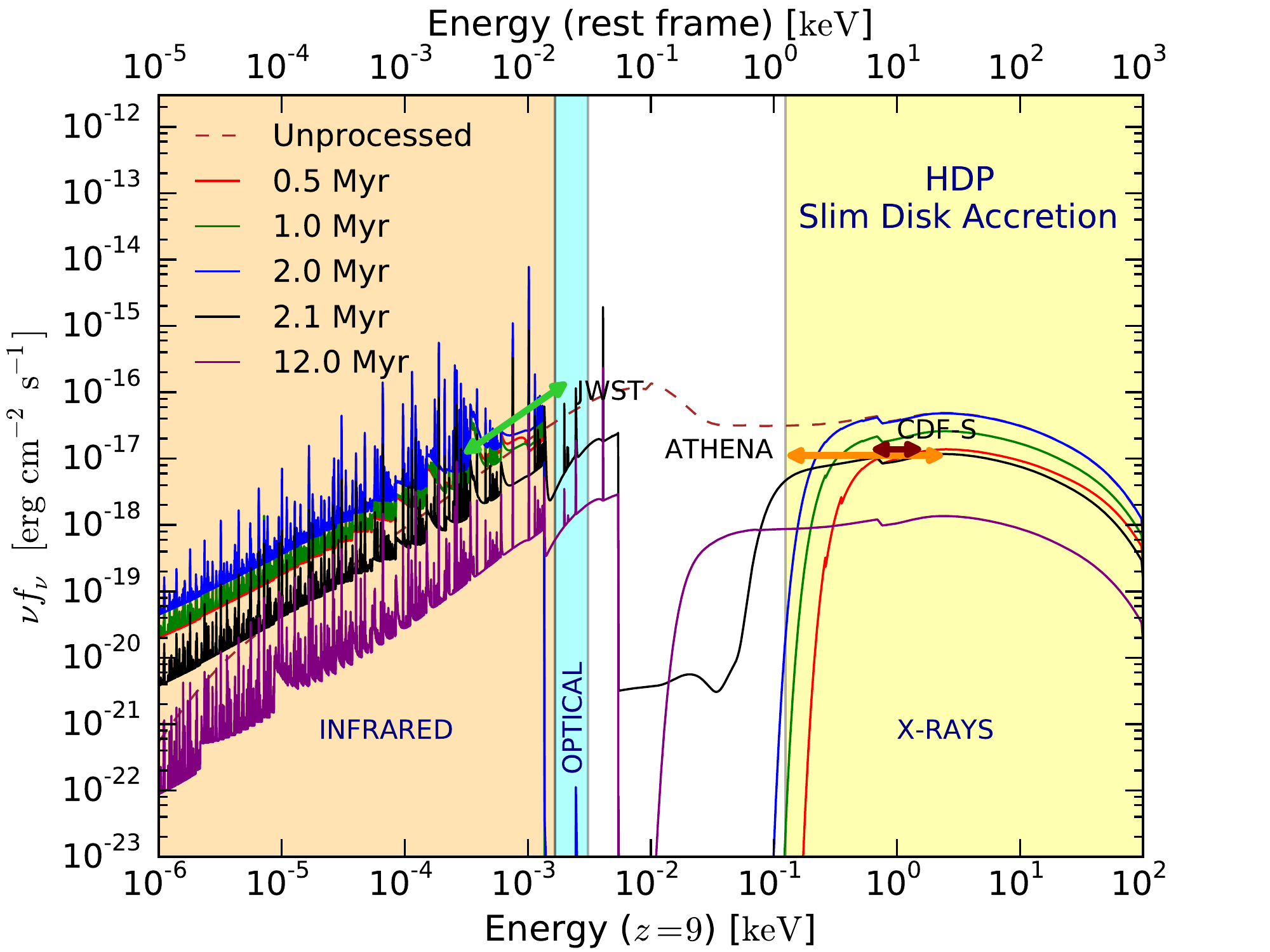}
\caption{Slim disk accretion - HDP case.}
\end{center}
\end{figure}


\bibliographystyle{mnras}
\bibliography{ms}

\label{lastpage}
\end{document}